\documentclass[prd,tightenlines,nofootinbib,superscriptaddress]{revtex4}
\usepackage{amsfonts,amsmath,amssymb,amsthm,bbm}

\usepackage{color,psfrag}
\usepackage[dvips]{graphicx}

\newcommand{\C}{{\mathbb C}}
\newcommand{\N}{{\mathbb N}}
\newcommand{\R}{{\mathbb R}}

\newcommand{\cI}{{\mathcal I}}

\newcommand{\cL}{{\mathcal L}}

\newcommand{\cM}{{\mathcal M}}

\newcommand{\SU}{\mathrm{SU}}
\newcommand{\SL}{\mathrm{SL}}
\newcommand{\SO}{\mathrm{SO}}
\newcommand{\Spin}{\mathrm{Spin}}

\newcommand{\be}{\begin{equation}}
\newcommand{\ee}{\end{equation}}
\newcommand{\bes}{\begin{eqnarray}}
\newcommand{\ees}{\end{eqnarray}}
\newcommand{\beq}{\begin{eqnarray}}
\newcommand{\eeq}{\end{eqnarray}}
\newcommand{\bea}{\begin{eqnarray}}
\newcommand{\eea}{\end{eqnarray}}
\newcommand{\nn}{\nonumber}

\newcommand{\Ref}[1]{(\ref{#1})}
\newcommand{\mat}[2]{\left(\begin{array}{#1}#2\end{array}\right)}

\newcommand{\su}{{\mathfrak su}}
\newcommand{\spin}{{\mathfrak spin}}

\newcommand{\la}{\langle}
\newcommand{\ra}{\rangle}

\newcommand{\tr}{{\mathrm Tr}}
\newcommand{\f}{\frac}

\def\arr{\rightarrow}

\def\vphi{\varphi}

\def\om{\omega}
\def\Om{\Omega}
\newcommand{\id}{\mathbb{I}}

\newcommand{\vJ}{\vec{J}}
\newcommand{\vK}{\vec{K}}
\def\hh{{\cal H}}
\def\Hs{{H_S}}
\def\Hp{{H_p}}
\def\tpsi{\tilde{\psi}}
\def\tphi{\tilde{\phi}}
\def\tvphi{\tilde{\vphi}}
\def\th{\tilde{h}}
\def\tk{\tilde{k}}
\def\tj{\tilde{j}}
\def\tbeta{\tilde{\beta}}
\def\trho{\tilde{\rho}}

\begin{document}

\title{Lifting SU(2) Spin Networks to Projected Spin Networks}

\author{{\bf Ma\"it\'e Dupuis}\email{maite.dupuis@ens-lyon.fr}}
\affiliation{Laboratoire de Physique, ENS Lyon, CNRS-UMR 5672, 46
All\'ee d'Italie, Lyon 69007, France}
\author{{\bf Etera R. Livine}\email{etera.livine@ens-lyon.fr}}
\affiliation{Laboratoire de Physique, ENS Lyon, CNRS-UMR 5672, 46
All\'ee d'Italie, Lyon 69007, France}

\date{\today}

\begin{abstract}

Projected spin network states are the canonical basis of quantum states of geometry for the most recent EPR-FK spinfoam models for quantum gravity. They are functionals of both the Lorentz connection and the time normal field. We analyze in details the map from these projected spin networks to the standard $\SU(2)$ spin networks of loop quantum gravity. We show that this map is not one-to-one and that the corresponding ambiguity is parameterized by the Immirzi parameter. We conclude with a comparison of the scalar products between projected spin networks and $\SU(2)$ spin network states.

\end{abstract}

\maketitle

\tableofcontents


\section*{Introduction}

The spinfoam framework is a proposal for a regularized path integral for quantum gravity. It was first constructed in order to provide us with a history formalism for Loop Quantum Gravity (LQG), thus defining dynamics and transition amplitudes between spin network states of quantum gravity.
However, most of the spinfoam models for 4d gravity have been constructed as discretized path integral for constrained BF field theories with the Lorentz group $\SL(2,\C)$ as gauge group. Their boundary are resulting $\SL(2,\C)$-invariant spin network states while the kinematical Hilbert space of Loop Quantum Gravity is spanned by $\SU(2)$ spin networks. This has been an essential discrepancy creating a gap between the original LQG theory and the developing spinfoam framework.

An early attempt to bridge between these two framework was proposed by Alexandrov and collaborators \cite{clqg,clqg0,clqg1,clqg2}\footnotemark.
\footnotetext{
Another early attempt was to define spinfoam models based on the self-dual Ashtekar connection instead of the Lorentz connection, thus directly using $\SU(2)$ spin networks (see e.g. \cite{michael}).
}
This canonical formalism hints towards a ``Covariant Loop Quantum Gravity" which uses {\it projected spin network states} introduced by one of the authors \cite{projected}. These projected spin networks are $\SL(2,\C)$-invariant states, which are functional of both the Lorentz connection and the time-normal field (partially defining the embedding of the canonical hypersurface in the 4d space-time manifold). From the spinfoam point of view, the first  explicitly constructed spinfoam model was the Barrett-Crane model \cite{bcl,bcl_carlo} and it was shown that it could be reformulated in term of the same projected spin network states \cite{projected,bcl_normal}.
However, the precise boundary states of the Barrett-Crane models were a very special case of projected spin network states, called simple spin networks, and they could not be put in one-to-one correspondence with $\SU(2)$ spin networks thus not allowing an easy translation of the Barrett-Crane spinfoam amplitudes to  LQG transition amplitudes.

This approach was given a second chance with the more recent EPRL-FK spinfoam models. Indeed, a new spinfoam model was proposed to address  the shortcomings of the Barrett-Crane models and a map between its boundary states and $\SU(2)$ spin networks was introduced thus finally hinting towards a direct and explicit relationship between spinfoams and loop gravity \cite{epr}. This EPR model was quickly generalized to the EPRL-FK models, which were constructed in both Euclidean and Lorentzian signature and taking into account non-trivial values of the Immirzi parameter \cite{eprl,fk,ls2,roberto}. These spinfoam models are based on a reformulation of the simplicity constraints involving the time-normal field. These simplicity constraints are the essential ingredient of the spinfoam program, turning topological BF theory into general relativity. Considering the non-trivial role of the time-normal field in the simplicity constraints \cite{epr,eprl,ls2,sf_valentin,sf_daniele,sf_jimmy} and the construction of the resulting spinfoam amplitudes, one can argue that it should not be considered as a mere mathematical and technique tool but considered as a relevant variable of the theory on the same footing than the Lorentz connection \cite{sf_sergei1,sf_sergei2}. This point of views leads to considering projected spin networks as the natural boundary states for the EPRL-FK spinfoam amplitudes.

Recently, the isomorphism between $\SU(2)$ spin networks and EPRL-FK boundary states defined as $\SL(2,\C)$ spin networks has been investigated in details \cite{eprl_jerzy}. We propose to revisit this correspondence using projected spin networks as EPRL-FK boundary states. Projected spin networks were already shown to have a mathematical structure very close to  $\SU(2)$ spin networks \cite{projected}. Here we pursue the line of research initiated in \cite{projected} and we investigate in details the correspondence between $\SU(2)$ spin networks and projected  spin networks. More particularly, we look at how to consistently map $\SU(2)$ spin networks onto projected  spin networks. Thus, instead of postulating one spinfoam model and  trying to map its boundary states on LQG's canonical states, we follow the reverse problematic: what are the various ways to map LQG's spin networks onto certain subspaces of projected spin networks. Each such space of projected spin networks could then be the Hilbert space of boundary states of a spinfoam model which would legitimately implement the dynamics and evolution of  $\SU(2)$ spin networks for Loop Quantum Gravity.

In a first section, we review the framework of projected spin networks and analyze in details their invariance properties. Then in the second section, we introduce a projection map from projected spin networks down to $\SU(2)$ spin networks and we investigate the inverse lift operators which would reversely map $\SU(2)$ spin networks up to the projected spin networks. We take particular care of the role of the Immirzi parameter and focus on the projected spin networks relevant to the EPRL-FK spinfoam models.

\section{A Review of Projected Spin Networks}

\subsection{Cylindrical Functions and Gauge Invariance}

Let us consider an arbitrary oriented graph $\Gamma$ with $E$ edges and $V$ vertices. We now look at the space of functions over $\SL(2,\C)^E\times \hh_+^V$. Here we have introduced the hyperboloid $\hh_+=\{x| x_\mu x^\mu=1,x^0>0\}$ made of unit time-like vectors in the Minkowski space $\R^{3,1}$ with signature $(+---)$. This hyperboloid is equivalently defined as the coset space $\hh_+\sim\SL(2,\C)/\SU(2)\sim\SO(3,1)/\SO(3)$.

Such functions are to be considered as functionals of the Lorentz connection and the time-normal field, living on the canonical hypersurface. They are called cylindrical in that they depend on these fields through only a finite number of degrees of freedom, more precisely the holonomies of the Lorentz connection along the edges of the graph $\Gamma$ and the values of the time-normal field at its vertices.

We further require that our functionals be invariant under the action of the Lorentz group:
\be
\vphi(G_e,x_v)=\vphi(\Lambda_{s(e)}G_e\Lambda_{t(e)}^{-1},\Lambda_v\vartriangleright x_v),\quad
\forall \Lambda_v\in\SL(2,\C)^{\times V},
\ee
where $G_e$ are the $\SL(2,\C)$ group elements and $x_v$ the 4-vectors in $\hh_+$. $s(e)$ and $t(e)$ are respectively the source and target vertices of the edge $e$.

The 4-vector $\Lambda_v\vartriangleright x_v$ is  obtained by acting  on $x_v$ by the $\SO(3,1)$ transformation corresponding to $\Lambda_v\in\SL(2,\C)$. The easiest way to write this action is to represent 4-vectors as 2$\times$2 Hermitian matrices:
\be
x\,\arr\,X=\mat{cc}{x_0+x_3 & x_1+ix_2 \\x_1-ix_2 & x_0-x_3},
\ee
with $\tr X=2x_0$ and $\det X=|x|^2$. Then $\SL(2,\C)$ group elements act by conjugation: $\Lambda\vartriangleright X \equiv\, \Lambda X \Lambda^\dag$. From there, we can act on the 4-vector $\om=(1,0,0,0)$, or equivalently on its corresponding matrix $\Om=\id$, to generate all elements in $\hh_+$:
\be
x= B\vartriangleright \om,\qquad
X=B\id B^\dag=B B^\dag,
\ee
for $B\in\SL(2,\C)$.
It is clear that this expression is invariant under the right $\SU(2)$ action $B\arr B h$ with $h\in\SU(2)$, since $h^\dag=h^{-1}$. This actually shows the fact that $\hh_+$ is the coset $\SL(2,\C)/\SU(2)$.
From these various representations, we can equivalently see our functionals as depending on 4-vectors, 2$\times$2 Hermitian matrices or $\SL(2,\C)$ group elements (with an extra $\SU(2)$ invariance), i.e respectively $\vphi(G_e,x_v)$ or $\vphi(G_e,X_v)$ or $\vphi(G_e,B_v)$.

\medskip

A first important remark on these Lorentz invariant functions is that they are entirely determined by their section at $x_v=\om$ for all $v$. Indeed, let us define this section:
\be
\phi(G_e)\equiv\vphi(G_e,x_v=\om).
\ee
Effectively, these functions still satisfy a remaining $\SU(2)$-invariance, inherited from the full $\SL(2,\C)$-invariance:
\be
\phi(G_e)=\phi(h_{s(e)}G_eh_{t(e)}^{-1}),\quad\forall h_v\in\SU(2)^{\times V}.
\ee
And we can reconstruct the full functional from that particular section:
\be
\vphi(G_e,x_v)=\vphi(G_e,B_vB_v^\dag)=\phi(B_{s(e)}^{-1}G_eB_{t(e)}).
\ee
The second remark is that if we integrate over the time-normals, then we recover the standard
$\SL(2,\C)$-invariant cylindrical functions, whose basis are $\SL(2,\C)$ spin networks.
More precisely, we define the group-averaged functional
\be
\vphi_g(G_e)=\int_{\hh_+^V} [dx_v]\, \vphi(G_e,x_v)=\int_{\SL(2,\C)^V} [d\Lambda_v] \, \vphi(G_e,\Lambda_v\vartriangleright\om),
\ee
where $[dx]$ is the translation-invariant measure on $\hh_+$ inherited from the Haar measure $[d\Lambda]$ on $\SL(2,\C)$. This new function satisfy a simple $\SL(2,C)$-invariance at the vertices:
\be
\vphi_g(G_e)=\vphi_g(\Lambda_{s(e)}G_e\Lambda_{t(e)}^{-1}),\quad
\forall \Lambda_v\in\SL(2,\C)^{\times V},
\ee

\medskip

Following \cite{projected}, the next step is to endow our space of cylindrical functions with a scalar product:
\be
\la \vphi | \vphi'\ra
\,\equiv\,
\int [dG_e]\, \overline{\vphi}(G_e,x_v)\vphi'(G_e,x_v).
\ee
Due to the $\SL(2,\C)$ gauge invariance satisfied by the functionals, it is easy to see that this definition holds for any arbitrary choice of time-normals $x_v$ as long as both functionals are evaluated on the same set of $x_v$'s of course. Therefore, this scalar product can be entirely computed by setting all time-normals to the origin $\om$:
\be
\la \vphi | \vphi'\ra
\,=\,
\int [dG_e]\, \overline{\phi}(G_e)\phi'(G_e).
\ee
We call the corresponding $L^2$ space of functions as the Hilbert space of projected cylindrical functionals on the graph $\Gamma$, following the terminology introduced in \cite{projected}, and we will simply write $H$ for it (leaving implicit the dependence on the underlying graph $\Gamma$, since our whole analysis does not involve changing graph).

The next step is to introduce the basis of $H$ given by the projected  spin networks. To this purpose, we need to recall a few facts about the unitary representations of the Lorentz group $\SL(2,\C)$.

\subsection{Quick Overview of $\SL(2,\C)$ Representations}

The Plancherel decomposition formula for $\SL(2,\C)$ states that $L^2$ functions with respect to the Haar measure on $\SL(2,\C)$ uniquely decompose in term of the matrix elements of the group element in the unitary irreducible  representations of $\SL(2,\C)$ of the principal series. Such irreducible representation (irreps) are labeled by a couple of numbers $(n,\rho)$, where $n\in\N/2$ is a half-integer and $\rho\in\R$ a real number.
There also exists a supplementary series of unitary irreps, labeled by a single real number bounded by 1 in modulus, but they do not enter the Plancherel decomposition.
Then the Plancherel formula for a function $f\in L^2(\SL(2,\C))$ reads:
\be
f(G)=\f1{8\pi^4}\sum_n\int \mu(n,\rho) d\rho\, \tr\,\left[F(n,\rho)D^{(n,\rho)}(G)\right],
\ee
where the Fourier components $F(n,\rho)$ are matrices in the Hilbert space of the representation $(n,\rho)$ and are obtained by the reverse formula:
\be
F(n,\rho)=\int dG\,f(G)D^{(n,\rho)}(G^{-1}).
\ee
The measure of integration over the representation labels $\mu(n,\rho) d\rho \,\equiv(n^2+\rho^2)d\rho$ is called by the Plancherel measure. This Plancherel decomposition relies on the fact that the matrix elements $D^{(n,\rho)}(G)$ form an orthogonal basis for the Hilbert space $L^2(\SL(2,\C))$.


It will be useful for later to have the explicit action of the Lorentz generators in each $(n, \rho)$ representation. The relevant basis for us is the $\SU(2)$ basis obtained by decomposing the $\SL(2,\C)$ representation into $\SU(2)$ irreducible representations. Indeed, one can show that the $(n, \rho)$ representation decomposes onto all $\SU(2)$ irreps with spin $j$ bounded below by the half-integer $n$, thus implying that the Hilbert space of the $(n, \rho)$ representation is the direct sum of the Hilbert spaces corresponding to these $\SU(2)$  irreps:
\be
R^{(n,\rho)}=\bigoplus_{j\in n+\N} V^j.
\ee
Let us point out that we have chosen the canonical $\SU(2)$ subgroup, which stabilizes the 4-vector $\om$ or equivalently the identity matrix $\Om=\id$ (as explained previously). Then we give the action of the $\su(2)$-rotation generators $\vJ$ and boost generators $\vK$ in the standard basis for $\SU(2)$-representations in term of the spin $j$ and the magnetic momentum $m$, diagonalizing the rotation operator $J^3$:
\bes
J^3\,|j, m\ra &=& m|j,m\ra,  \label{actionJ}\\
J^+\,|j,m\ra&=& \sqrt{(j-m)(j+m+1)}\,|j,m+1\ra, \nn \\
J^-\,|j,m\ra &=&\sqrt{(j+m)(j-m+1)}\,|j,m-1\ra, \nn \\
K^3\,|j,m\ra&=&- \alpha_j \sqrt{j^2-m^2}\, |j-1,m\ra -\beta_jm \, |j,m\ra + \alpha_{j+1} \sqrt{(j+1)^2-m^2}\, |j+1,m\ra, \label{actionK}\\
K^+\,|j,m\ra&=&-\alpha_j \sqrt{(j-m)(j-m-1)}\, |j-1,m+1\ra -\beta_j \sqrt{(j-m)(j+m+1)} \, |j, m+1\ra \nn \\
&&  -\alpha_{j+1} \sqrt{(j+m+1)(j+m+2)}\,|j+1,m+1\ra, \nn \\
K^-\,|j,m\ra&= & \alpha_j \sqrt{(j+m)(j+m-1)}\, |j-1,m-1\ra- \beta_j \sqrt{(j+m)(j-m+1)} \, |j,m-1 \ra \nn \\
&& +\alpha_{j+1} \sqrt{(j-m+1)(j-m+2)}\,|j+1,m-1\ra , \nn
\ees
where the coefficients defining the action of the boost generators are given by:
\be
\label{actionbeta}
\beta_j=\f{ n\rho}{j(j+1)},
\qquad \alpha_j= \f{i}{j} \sqrt{\f{(j^2-n^2)(j^2+\rho^2)}{4j^2-1}}.
\ee
It is straightforward to check that this postulated action satisfied as expected the $\SL(2,\C)$ commutation relations~\footnotemark.
\footnotetext{
The  commutation relation of the $\SL(2,\C)$ Lie algebra are:
\bes
&&[J^+,J^3]=-J^+, \quad[J^-, J^3]=J^-, \quad [J^+, J^-]=2J^3 \nn \\
&& [J^+,K^+]=[J^-, K^-]= [J^3, K^3]=0, \quad [J^+, K^-]=-[J^-, K^+]=2K^3, \nn \\
&&[J^+, K^3]=-K^+, \quad [J^-,K^3]=K^-, \quad [K^+, J^3]=-K^+, \quad [K^-, J^3]=K^-, \nn\\
&&[K^+,K^3]=J^+, \quad[K^-, K^3]=-J^-, \quad [K^+, K^-]=-2J^3. \nn
\ees
}
Moreover, since the coefficient $\alpha_n=0$ vanishes for $j=n$, it is also clear that the truncation to spins $j\ge n$ is self-consistent. On the other hand, it is obvious that the coefficients $\alpha_j$ for $j>n$ will never vanish, thus there is no upper bound on the spin $j$. This is consistent with the fact that a unitary representation of $\SL(2,\C)$ necessarily has an infinite dimension.

From this action, we can check that the $\SU(2)$ Casimir operator has the usual value $\vJ^2=j(j+1)$. We can also compute the values of the two Casimir operators of $\SL(2,\C)$:
\be
C_1=\vec{K}^2-\vJ^2=\rho^2-n^2+1,
\qquad
C_2= \vJ \cdot \vec{K}=2n \rho.
\ee

Finally, we introduce the characters of the $\SL(2,\C)$ representations, $\Theta^{(n,\rho)}(G)\,\equiv\,\tr\,D^{(n,\rho)}(G)$. It is easy to evaluate it on $\SU(2)$ group elements since we know the decomposition of the $\SL(2,\C)$ representation into $\SU(2)$ representations~\footnotemark:
\be
\forall g\in\SU(2),\,
\Theta^{(n,\rho)}(g)
\,=\,
\sum_{j\in n+\N} \chi^j(g)
\,=\,
\sum_{j\in n+\N} \f{\sin(2j+1)\theta}{\sin\theta}
\,=\,
\f{\cos 2n\theta}{2\sin^2\theta},
\ee
where $\theta$ is the class angle of the group element $g$, i.e meaning that $g$ is conjugate to the diagonal matrix with entries $[e^{i\theta},e^{-i\theta}]$.
\footnotetext{
The character formula is straightforwardly generalizable to the whole $\SL(2,\C)$ group. Indeed, all group elements are conjugated to a diagonal matrix. Then we can evaluate the character on such matrices (see e.g. \cite{noncompact}):
$$
\Theta_{(n,\rho)}
\mat{cc}{e^{\lambda+i\theta} & 0 \\ 0&e^{-\lambda-i\theta}}
\,=\,
\f{e^{i\rho\lambda}e^{i2n\theta}+e^{-i\rho\lambda}e^{-i2n\theta}}{|e^{\lambda+i\theta}-e^{-\lambda-i\theta}|^2}.
$$
}

Now that we have quickly reviewed these basic facts on $\SL(2,\C)$ unitary representations and the Plancherel decomposition, we are ready to introduce the basis of projected spin networks for our Hilbert space $H$ of Lorentz invariant cylindrical functions.

\subsection{The Basis of Projected Spin Networks}

Our goal is to build a basis of the Hilbert space $H$ of Lorentz invariant functions $\vphi(G_e,x_v)$. Following the original work \cite{projected}, we start with the section $\phi(G_e)=\vphi(G_e,\om)$, which fully determines the whole function $\vphi(G_e,x_v)$. We apply the Plancherel decomposition formula to $\phi(G_e)$, thus attaching an irrep $(n_e,\rho_e)$ and the corresponding matrix $D^{(n_e,\rho_e)}(G_e)$ to each edge $e$ of the graph. Then we glue these matrices at each vertex $v$ of the graph with vectors in the tensor product of the irreps attached to the incoming/outgoing edges. These tensors are not chosen entirely arbitrarily since the functions $\phi(G_e)$ are required to be $\SU(2)$-invariant at each vertex.

The final result of this procedure are the {\it projected spin networks}. A projected spin network on the graph $\Gamma$ is defined by the choice of a $\SL(2,\C)$ irrep $\cI_e=(n_e,\rho_e)$ for each edge, a choice of couple of $\SU(2)$ irrep $(j_e^s,j_e^t)$ attached to the source and target vertices of  each edge, and finally a $\SU(2)$-intertwiner (or equivalently $\SU(2)$-invariant tensor, or a singlet state in layman terminology) $i_v$ for each vertex $v$. The intertwiner $i_v$ lives in the tensor product of the $\SU(2)$ irreps coming in and going out the vertex $v$, or more precisely:
$$
i_v\,:\,
\bigotimes_{e|s(e)=v} V^{j_e^s}
\,\longrightarrow\,
\bigotimes_{e|t(e)=v} V^{j_e^t}.
$$
Then the functions is defined as:
\be
\vphi_{\cI_e,j_e^{s,t},i_v}(G_e,x_v)
\,\equiv\,
\tr
\prod_e \la \cI_e,j_e^s,m_e^s|B_{s(e)}^{-1}G_e B_{t(e)}| \cI_e,j_e^t,m_e^t\ra
\,
\prod_v \la \otimes_{e|t(e)=v}\,\cI_e,j_e^t,m_e^t| i_v |\otimes_{e|s(e)=v}\cI_e,j_e^s,m_e^s\ra.
\ee
The trace is taken over the $\SU(2)$ representations i.e it amounts to summing over the basis labels $m_e^{s,t}$. We must require that the choice of spins $j_e^{s,t}$ be compatible with the choice of the $\SL(2,\C)$ irreps $\cI_e=(n_e,\rho_e)$, i.e that $j_e^{s,t}\ge n_e$, else the projected spin network functional would simply vanish.

\begin{figure}[h]
\psfrag{xv}{$x_v$}
\psfrag{iv}{$i_v$}
\psfrag{Ie}{$(n_e,\rho_e)$}
\psfrag{js}{$j_e^s$}
\psfrag{jt}{$j_e^t$}
\psfrag{Ge}{$G_e$}
\begin{center}
\includegraphics[height=30mm]{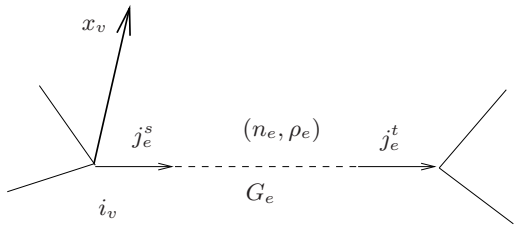}
\caption{An edge of a projected spin network}
\end{center}
\end{figure}

First, to check that this function is well-defined, one must make sure that its definition is invariant under the right $\SU(2)$-action on the group elements $B_v$. It is actually the requirement of having $\SU(2)$-invariant intertwiners $i_v$ which ensures that the expression above is correctly invariant under the transformation $B_v\arr B_v h_v$ for all $h_v\in\SU(2)^{\times V}$.

Then, we would like to check that these projected spin networks are properly $\SL(2,\C)$-invariant. The Lorentz action at the vertices reads as:
$$
\left|
\begin{array}{l}
G_e \\
x_v \\
B_v
\end{array}
\right.
\,\longrightarrow\,
\left|
\begin{array}{l}
\Lambda_{s(e)}G_e\Lambda_{t(e)}^{-1} \\
\Lambda_v\vartriangleright x_v \\
\Lambda_v B_v
\end{array}
\right. \,.
$$
It is clear that the functions defined above are invariant under such transformations.

Finally, the Plancherel decomposition formula ensures that these projected spin network functionals cover the whole Hilbert space $H$ and provide us with an orthonormal basis. Indeed we can compute the scalar product between two such spin networks:
\be
\la \vphi_{\cI_e,j_e^{s,t},i_v}| \vphi_{\tilde{\cI}_e,\tilde{j}_e^{s,t},\tilde{i}_v}\ra
\,=\,
\int [dG_e]\,
\overline{\phi}_{\cI_e,j_e^{s,t},i_v}(G_e)
\phi_{\tilde{\cI}_e,\tilde{j}_e^{s,t},\tilde{i}_v}(G_e)
\,=\,
\prod_e\f{\delta_{n_e,\tilde{n}_e}\delta(\rho_e-\tilde{\rho}_e)}
{\mu(n_e,\rho_e)}
\,\delta_{j_e^{s,t},\tilde{j}_e^{s,t}}
\,\prod_v \la i_v|\tilde{i}_v\ra.
\label{scalarproj}
\ee
Thus a choice of orthonormal basis is given by a choice of an orthonormal basis of  $\SU(2)$ intertwiners, just as for the standard $\SU(2)$ spin networks of Loop Quantum Gravity.

\section{Back and Forth Between Projected and $\SU(2)$ Spin Networks}

We have reviewed in the previous section the projected spin networks, which are the natural boundary states for Spin Foam models. Our goal is to compare them with the $\SU(2)$ spin network basis of Loop Quantum Gravity. As we have seen, the projected spin networks are Lorentz-invariant functionals of the $\SL(2,\C)$ connection and of the time-normal field. Nevertheless, as soon as we fix the value of the time-normal field (at the vertices of the graph used to construct the spin network), they are only required to satisfy an effective $\SU(2)$ invariance and thus they are built using $\SU(2)$-intertwiners and not $\SL(2,\C)$-intertwiners. Since $\SU(2)$ spin networks are also built from $\SU(2)$-intertwiners, this hints towards a direct path between the two sets of states. From this perspective, projected spin networks seems to be extensions of $\SU(2)$ spin networks, allowing to evaluate them on the whole Lorentz group $\SL(2,\C)$ and not only on the $\SU(2)$ subgroup.

\subsection{Projecting down to $\SU(2)$ Spin Networks}

Let us start by reminding the definition of $\SU(2)$ cylindrical functions on the graph $\Gamma$. They are functions of $E$ group elements in $\SU(2)$ living on the edges of the graph and satisfying a $\SU(2)$ invariance at every vertex:
\be
\psi(g_e)=\psi(h_{s(e)}g_e h_{t(e)}^{-1}),\quad\forall h_v\in\SU(2)^{\times V}.
\ee
The natural scalar product on this space of functions is:
\be
\la\psi|\psi'\ra_{\SU(2)}
\,=\,
\int_{\SU(2)}[dg_e]\,
\overline{\psi}(g_e)\psi'(g_e),
\ee
where $dg$ is the Haar measure on the $\SU(2)$ Lie group. Let us call $\Hs$ the $L^2$ space of such $\SU(2)$ invariant cylindrical functions. Then this Hilbert space $\Hs$ is spanned by the usual spin network states. A spin network is labeled by a set of spins $j_e$ for each edge and $\SU(2)$-intertwiners $i_v$ for every vertex. Then we define:
\be
\psi_{j_e,i_v}(g_e)
\,\equiv\,
\tr
\prod_e \la j_e,m_e^s|g_e| j_e,m_e^t\ra
\,
\prod_v \la \otimes_{e|t(e)=v}j_e,m_e^t| i_v |\otimes_{e|s(e)=v}j_e,m_e^s\ra,
\ee
which simply amounts to contracting the Wigner matrices $D^j_{m^sm^t}(g)=\la j,m^s|g| j,m^t\ra$ along every edge $e$ with the intertwiners sitting at the vertices. We point out that this definition is almost the same as the one of projected spin networks: the difference is that we evaluate projected spin networks on the whole $\SL(2,\C)$ group and this requires the choice of an extra  $\SL(2,\C)$ irrep $\cI_e$ for each edge of the graph.

The scalar product between two such $\SU(2)$ spin networks is easily computed:
\be
\la\psi_{j_e,i_v}|\psi_{\tj_e,\tilde{i}_v}\ra_{\SU(2)}
\,=\,
\prod_e\f{\delta_{j_e,\tilde{j}_e}}{d_{j_e}}
\,
\,\prod_v \la i_v|\tilde{i}_v\ra,
\label{scalarsu}
\ee
where we remind that $d_j=(2j+1)$ is the dimension of the $\SU(2)$-irrep of spin $j$.

\medskip

Since the projected cylindrical functions and the $\SU(2)$ cylindrical functions share the same $\SU(2)$ invariance, it is natural to introduce the following projection:
\be
\begin{array}{llcl}
\cM:\,&H&\rightarrow&H_S \\
&\vphi(G_e,x_v)&\mapsto&
\psi(g_e)=\vphi(g_e,\om)=\phi(g_e),
\end{array}
\ee
which is simply the restriction of the projected cylindrical function to the $\SU(2)$ subgroup.
Considering the invariance property of the function $\vphi$ and its section $\phi$ at $x_v=\om,\,\forall v$, the map $\cM$ is well-defined and the resulting function $\psi$ is correctly $\SU(2)$-invariant as wanted.

It is straightforward to compute the image of the projected spin network by the map $\cM$. First, considering the case of functions with $j_e^s\ne j_e^t$, the corresponding $\SU(2)$ function vanishes:
\be
\forall j_e^s\ne j_e^t,\quad \cM\vphi_{\cI_e,j_e^{s,t},i_v}\,=\,0,
\ee
since a $\SU(2)$ group element could never trigger a transition between two different $\SU(2)$ irreps (by definition). On the other hand, now assuming that the two spins are equal for all edges so that we can drop the index $s,t$, $j_e^s=j_e^t=j_e$, then the image of the corresponding projected spin network is as expected simply a $\SU(2)$ spin network:
\be
\forall j_e^s=j_e^t=j_e,\quad
\cM\vphi_{\cI_e,j_e,i_v}\,=\,\psi_{j_e,i_v},
\ee
as long as the spin $j_e$ is compatible with the $\SL(2,\C)$ irrep, i.e $j_e\ge n_e$ (or more exactly $j_e\in n_e+\N$).

\medskip

In the next sections, we investigate the inverse map(s) to $\cM$, that is how to lift $\SU(2)$ cylindrical functions to functions on the whole Lorentz group $\SL(2,\C)$. Understanding in details how this lifting is achieved is crucial to the construction of the EPR-FK class of spin foam models and their interpretation as an ansatz for the dynamics of Loop Quantum Gravity.

In the following, we will focus on projected spin networks satisfying the ``matching" constraints $j_e^s=j_e^t$. We call $\Hp$ the Hilbert spanned by these ``proper" projected spin network functionals (whose evaluation on the $\SU(2)$ subgroup does not trivially vanish). As seen from the last equation above, inverting the map $\cM$ would more or less simply amount to choosing a $\SL(2,\C)$ irrep $\cI_e$ into which to embed the $\SU(2)$ irrep $j_e$. We analyze this in details below.

\subsection{Lifting back Spin Networks}
\label{lifting}

Starting with a $\SU(2)$ cylindrical function $\psi(g_e)$ invariant under the $\SU(2)$ action at every vertex, the goal is to construct a Lorentz invariant extension for it. Following the insight of the previous section, the simplest way to proceed would be to decompose the function $\psi$ in $\SU(2)$ irrep $j_e$ and to choose a $\SL(2,\C)$ irrep for every spin. At the level of the groups, these operations are done through convolutions with $\SU(2)$ and $\SL(2,\C)$ characters.

More precisely, starting with $\psi(g_e)$, we construct the following projected cylindrical function:
\be
\vphi(G_e,B_v)
\,\equiv\,
\sum_{\{j_e\}}\Delta_{j_e}\int_{\SU(2)} [dh_edk_e]\,
\psi(k_e)\,\chi^{j_e}(h_ek_e)\,\Theta^{\cI_e}(B_{s(e)}^{-1}G_eB_{t(e)}h_e).
\label{lift}
\ee
$\Delta_{j}$ is a weight depending on the spin $j$ that we will uniquely fix below by requiring that $\cM\vphi=\psi$ or more explicitly $\vphi(g_e,\id)=\psi(g_e)$. The label $\cI_e$ is an arbitrary function of the spin $j_e$ and it does not need to be the same for all the edges $e$ of the graph. The only constraint is that the $\SU(2)$-irrep $j_e$ needs to be in the $\SL(2,\C)$-irrep $\cI_e$, i.e we require that $n_e\le j_e$ always (more exactly, $j_e\in n_e+\N$).

First, we check that the constructed function is invariant under $\SU(2)$ shifts $B_v\arr B_vh_v$. This is true thanks to the $\SU(2)$ invariance of the original function $\psi$.
Then, we easily see that this function is invariant under Lorentz transformations acting simultaneously on both $G_e$ and $B_v$.
Finally, we would like to ensure that $\vphi$ is a proper lifting of $\psi$, i.e that $\cM\vphi=\psi$. To check this, we compute straightforwardly the value of $\vphi$ for $G_e=g_e\in\SU(2)$ and $B_v=\id$:
\be
\vphi(g_e,\id)
\,\equiv\,
\sum_{\{j_e\}}\Delta_{j_e}\int_{\SU(2)} [dh_edk_e]\,
\psi(k_e)\,\chi^{j_e}(h_ek_e)\,\Theta^{\cI_e}(g_eh_e).
\ee
As we reviewed earlier, we can express the $\SL(2,\C)$-character in term of the $\SU(2)$ characters when evaluated on $\SU(2)$ group elements:
$$
\Theta^{(n_e,\rho_e)}(g_eh_e)
\,=\,
\sum_{l_e\in n_e+\N}\chi^{l_e}(g_eh_e).
$$
We can then proceed to the integration over $h_e$ using the known convolution formula\footnotemark{ } for $\SU(2)$-characters :
\footnotetext{
The convolution formula for $\SU(2)$-characters is:
$$
\int_{\SU(2)} dh\,\chi^{j}(hk)\chi^{l}(gh)=\,\f{\delta_{j,l}}{d_j}\,\chi^j(gk^{-1}),
$$
where $d_j=(2j+1)$ is the dimension of the $\SU(2)$-irrep of spin $j$. This follows from the orthonormality of matrix elements with respect to the Haar measure. When $g=k$ in particular, we recover the usual orthonormalization condition for characters $\int \chi^j\chi^l=\delta_{j,l}$.
}
$$
\vphi(g_e,\id)
\,=\,
\int_{\SU(2)} [dk_e]\,
\psi(k_e)\,\prod_e\sum_{j_e}\f{\Delta_{j_e}}{d_{j_e}}\,\chi^{j_e}(g_ek_e^{-1})
\,=\,
\psi(g_e),
$$
as long as we fix the weights $\Delta_j\,\equiv\, d_j^2=(2j+1)^2$ in order to recover the $\delta$-distribution, $\sum_j d_j\chi^j(gk^{-1})=\delta(gk^{-1})$.

\medskip

Finally, we have checked that our formula \Ref{lift} correctly defines a lift of $\SU(2)$ cylindrical functions to Lorentz-invariant projected cylindrical functions and properly inverses the projection map $\cM$. The parameters of this lifting are a choice of $\cI_e$ irrep label for each spin $j_e$ on each edge $e$. There have been two typical choices for this parameter in the spin foam literature:
\begin{itemize}

\item {\bf The Barrett-Crane ansatz}: $n_e=0$ for all spins $j_e$ on all edges \\

This restricts to irreps of the type $(0,\rho)$  used in the (Lorentzian) Barrett-Crane model \cite{bcl,bcl_carlo}. Let us emphasize that the label of the $\SL(2,\C)$ $n_e$ is not the spin $j_e$, which can still vary freely. If we further fix $j_e=0$, then we recover the simple spin networks usually used as boundary states of the Barrett-Crane model. Nevertheless, our analysis here suggests that we should {\it not} proceed to such a restriction and we would have a Hilbert space of projected spin networks for the BC model which would be isomorphic to the space of $\SU(2)$ spin networks. This interpretation of the BC model in term of projected spin networks and time-normals was already pushed forward in \cite{projected,clqg1,bcl_normal}. In particular, in \cite{bcl_normal}, it was speculated that spins $j_e\ne 0$ would correspond to particle insertions in the Barrett-Crane model, but we will not pursue in this direction.

\item {\bf The EPRL-FK ansatz}: $n_e=j_e$ for all spins $j_e$ on all edges \\

This is the condition to build $\SL(2,\C)$ coherent states  used in the construction of the Lorentzian spinfoam models of the EPRL-FK type \cite{eprl,fk,roberto}.
We will study this case in details in the next section, and see how the Immirzi parameter enters our definition of the inverse lift.

\end{itemize}

\subsection{Simplicity Constraints and the Immirzi Paramater}

\subsubsection{Weak Constraints}

Following the approach used for constructing the EPRL-FK spinfoam models, we look at weak constraints that are satisfied by the projected spin network states \cite{epr,eprl,fk,ls2}.
More precisely, we compare the matrix elements of the $\SU(2)$ rotation generators $\vJ$ and of the boost generators $\vK$ \cite{epr,eprl,eprl_ding}. The simplicity constraints amounts to requiring that the matrix elements of these two operators are the same up to a global factor, which would be identified as the Immirzi parameter.

We start with $\SU(2)$ spin network states $\psi$ and $\tpsi$, which we lift to projected spin networks $\vphi$ and $\tvphi$ using the same mapping i.e the same choice of $\SL(2,\C)$ irreps.
Then considering a fixed edge $e$, let us start by looking at the matrix elements of the left action of the boost generators $\vK_e$ on these projected spin networks:
\bes
\la \vphi | \vK_e^{(L)} | \tvphi\ra
&\equiv&
\int[dG_e]\,
\overline{\phi}(G_e)\vK_e\vartriangleright_L\tphi(G_e)\\
&=&
\int[dG_e][dh_ed\th_edk_ed\tk_e]\,
\overline{\psi}(k_e)\tphi(\tk_e)\,
\prod_e\sum_{j_e} d_{j_e}^2d_{\tj_e}^2\chi^{j_e}(h_ek_e)\chi^{\tj_e}(\th_e\tk_e)
\overline{\Theta^{(n_e,\rho_e)}}(G_eh_e)
\Theta^{(n_e,\rho_e)}(\vK_e G_e\th_e). \nn
\ees
The integral over the $\SL(2,\C)$ group elements $G_e$ can be done using the orthonormality of the $\SL(2,\C)$ matrix elements with respect to the Haar measure and give $\Theta^{(n_e,\rho_e)}(\vK_e h_e^{-1}\th_e)$ up to a measure factor depending solely on $(n,\rho)$.
Let us have a closer look at this term:
$$
\Theta^{(n_e,\rho_e)}(\vK_e h_e^{-1}\th_e)
\,=\,
\sum_{l_e,m_e} \la (n_e,\rho_e) l_e m_e| \vK_e h_e^{-1}\th_e |(n_e,\rho_e) l_e m_e\ra.
$$
First, the group variable $h_e^{-1}\th_e$ is in the $\SU(2)$ subgroup and therefore doesn't change the spin $l_e$. Thus only the matrix elements of the boost generators $\vK_e$ in the $\SU(2)$-irrep of spin $l_e$ matter. Next, due to the integration over $h_e$ and the insertion of the character $\chi^{j_e}(h_ek_e)$, only the component $l_e=j_e$ enters the calculation of the expectation value above. Similarly, the integration over $\th_e$ and the insertion of the character $\chi^{\tj_e}(\th_e\tk_e)$ forces $l_e=\tj_e=j_e$. Finally, we refer to the explicit action of the boost and rotation generators in a $(n,\rho)$-irrep given in \Ref{actionJ} and \Ref{actionK},
\be
\forall\, l,m,m',\quad
\la (n,\rho) l,m |\vK|(n,\rho) l,m' \ra
\,=\,
\beta_j^{(n,\rho)}\la (n,\rho) l,m |\vJ|(n,\rho) l,m' \ra,
\ee
where the coefficient $\beta_j$ is given in \Ref{actionbeta}. This was already noticed in \cite{clqg1,eprl,eprl_ding}. We would like to use this fact in order to relate the values of the expectation values $\la \vphi | \vK_e^{(L)} | \tvphi\ra$ and $\la \vphi | \vJ_e^{(L)} | \tvphi\ra$. The obvious issue is that $\beta_{j_e}^{(n_e,\rho_e)}$ depends on $j_e$ and the precise choice of embedding $(n_e,\rho_e)$ chosen for each value of $j_e$.

\medskip

Considering the Barrett-Crane ansatz $n_e=0$ for all values of $j_e$, we get the trivial value of the proportionality coefficients, $\beta_{j_e}^{(0,\rho_e)}=0$. This leads to the identity:
\be
\textrm{{\bf Barrett-Crane ansatz} }n_e=0
\quad\Rightarrow\quad
\la \vphi | \vK_e^{(L)} | \tvphi\ra =0.
\ee
We do not consider this ansatz particularly useful, but at least worth mentioning considering the attention that the Barrett-Crane model has received over the past decade.

\medskip

The case of the EPRL-FK ansatz is much more interesting. We choose the maximal value for the label of the $\SL(2,\C)$ irrep, $n_e=j_e$. Then we would like to fix the value of the coefficients $\beta_{j_e}$ to a fixed value $\beta_e$ which does not depend on the value of the spin $j_e$ but only on the considered edge $e$. This leads to a unique solution for $\rho_e$ as a function of the spin $j_e$:
\be
n_e(j_e)=j_e,\quad
\rho_e(j_e)=\beta_e(j_e+1),\quad
\Rightarrow\quad
\beta_{j_e}^{(n_e,\rho_e)}=\f{n_e\rho_e}{j_e(j_e+1)}\,=\,\beta_e.
\ee
This leads to the final equality:
\be
\textrm{{\bf EPRL-FK ansatz} } (n_e,\rho_e)=(j_e,\beta_e(j_e+1))
\quad\Rightarrow\quad
\la \vphi | \vK_e^{(L)} | \tvphi\ra
\,=\,
\beta_e\,\la \vphi | \vJ_e^{(L)} | \tvphi\ra.
\ee
The same equality holds if considering  the right action of the boost and rotation generators.
This is exactly the (linear) simplicity constraints that are imposed in the EPRL-FK spinfoam model with Immirzi parameter $\beta_e$.
Let us underline that we do not need to choose the same proportionality coefficient $\beta_e$ for all edges $e$. This is what is usually assumed in the EPRL-FK spinfoam model. However, in our framework, we are free to choose a different value $\beta_e$ for each edge of the graph, i.e a different value of the Immirzi parameter along the edges of the projected spin networks. This makes it more like an Immirzi field than an Immirzi parameter.

Finally, we introduce the precise lift inverting the projection map $\cM$ in the EPRL-FK ansatz. This lift is parameterized by a choice of coefficients $\{\beta_e\}\in\R^E$ for all edges of the graph. Then we define:
\be
\begin{array}{llcl}
\cL_{\{\beta_e\}}:
\,&H_S&\rightarrow&H_p \\
&\psi(g_e)&\mapsto&
\displaystyle{\vphi(G_e,B_v)=
\int_{\SU(2)} [dh_edk_e]\,
\psi(k_e)\,\sum_{j_e}d_{j_e}^2\,
\chi^{j_e}(h_ek_e)\,\Theta^{(j_e,\beta_e(j_e+1))}(B_{s(e)}^{-1}G_eB_{t(e)}h_e).}
\end{array}
\ee
As already shown in section \ref{lifting}, this provides us with a proper inverse for the map $\cM$:
\be
\forall \{\beta_e\},\quad
\forall \psi\in H,\quad
\cM\cL_{\{\beta_e\}}\psi\,=\,\psi.
\ee

\medskip

We can even go further by noticing by all possible values for $(n_e,\rho_e)\in\N/2\times\R$ are reached as $j_e$ and $\beta_e$ vary respectively in $\N/2$ and $\R$. Indeed, we can inverse the relations given above to get:
\be
j_e=n_e,\quad
\beta_e=\f{\rho_e}{j_e+1}.
\ee
This means that we can use the maps $\cL_{\{\beta_e\}}$ to obtain a full foliation of the Hilbert of (proper) projected spin network:
\be
H_p\,=\,
\bigoplus_{\{\beta_e\}\in\R^E}\,\cL_{\{\beta_e\}}H.
\ee
In words, this means that choosing arbitrary values of the Immirzi parameter $\beta_e$ for each edge of the graph, we will cover the whole space of proper projected spin networks by applying the lifting map $\cL_{\{\beta_e\}}$ to the standard $\SU(2)$ spin networks. We underline that we are restricted to {\it proper} projected spin networks since we always require that $j_e^s=j_e^t$ on all edges of the graph.

From the point of view of Loop Quantum Gravity's dynamics, we believe that the dynamical LQG operators would act on the Hilbert space $H$ of  standard $\SU(2)$ spin networks. This hints towards considering each subspace $\cL_{\{\beta_e\}}H$ of projected spin networks as {\it super-selection sectors} for the dynamics. A spinfoam model would then work in a given $\cL_{\{\beta_e\}}H$ subspace with all the parameters $\beta_e$ fixed, and would not mix these different sectors. Since spinfoam models are usually built for arbitrary graphs $\Gamma$, the simplest restriction would be to require that the Immirzi parameter be fixed and the same for all edges on all graphs, i.e $\beta_e=\beta,\,\forall e,\Gamma$. Then we recover the boundary states for the usual (Lorentzian) EPRL-FK spinfoam models with fixed Immirzi parameter.

Nevertheless, our framework leaves us the freedom of attributing a different value of the Immirzi parameter for each edge of the graph. Let us speculate on the possibility that the Immirzi parameter provides us with a (length/area) scale which we would vary when coarse-graining or renormalizing LQG's transition amplitudes and dynamics. Then our framework for boundary states would allow to coarse-grain various regions of space independently.

\subsubsection{Strong Constraints}

From the perspective of the construction of spinfoam models, the weak constraints can be translated to strong constraints in the spirit of ``master constraints". The logic is to replace the weak constraints  $\la \vphi|\vK_e-\beta_e\vJ_e|\tvphi\ra=0$ by strong constraints using the $\SU(2)$ and $\SL(2,\C)$ Casimir operators \cite{epr,eprl}.

Considering the EPRL-FK ansatz, $n(j)=j$ and $\rho(j)=\beta\,(j+1)$, we can easily express the values of the $\SL(2,\C)$ Casimir operators in term of the $\SU(2)$ Casimir operator:
\be
\begin{array}{lcl}
C_2= &\vJ\cdot\vK & = 2n\rho=2\beta j(j+1)=2\beta \vJ^2 \\
C_1= &\vK^2 -\vJ^2 & = \rho^2-n^2+1=(\beta^2-1)j(j+1)+(\beta^2+1)(j+1)=
(\beta^2-1)\vJ^2+(\beta^2+1)(\sqrt{\vJ^2+\f14}+\f12).
\end{array}
\ee
The expression of the second quadratic Casimir looks much simpler and it is straightforward to check that the explicit definition that the projected spin networks $\vphi=\cL_{\{\beta_e\}}\psi$ indeed satisfy strong (simplicity) constraints:
\be
\forall \vphi=\cL_{\{\beta_e\}}\psi,\quad
\left(\vJ_e\cdot\vK_e\,-2\beta_e\vJ_e^2\right)\,\phi=0.
\ee
Here, it does not matter whether we consider the left or right action of the boost and rotation operators as long as we take them all as acting on the same side of the group variable $G_e$. Moreover, we wrote the constraint as acting on the section $\phi(G_e)=\vphi(G_e,\om)$. This constraint can be rotated by the suitable Lorentz transformations to apply it on the whole function $\vphi(G_e,B_v)$.

As long as we require by hand that $n_e=j_e$, this strong constraint is sufficient to impose that $\rho_e=\beta_e\,(j_e+1)$. However, in order to impose $n_e=j_e$ through an operator constraint as well, we need to impose the other constraint involving the first Casimir operator. The drawback is that this constraint involve a rather ugly ``quantum correction" term in $\sqrt{\vJ^2}$ operator, which is nevertheless necessary if we want an exact constraint at the quantum level.

\subsection{Comparing $\SU(2)$ and $\SL(2,\C)$ Scalar Products}

Since we have constructed a map between $\SU(2)$ spin networks and projected spin networks, it is natural to wonder if these lifts are unitary and preserve the scalar products. It is straightforward to see that this is a priori not the case. Indeed, considering two projected cylindrical functions, $\vphi$ and $\tvphi$, and their projections $\psi=\cM\vphi,\tphi=\cM\tvphi$, the scalar products are best expressed in term of the sections $\phi,\tphi$:
\bes
\la \vphi |\tvphi\ra &=&\int_{\SL(2,\C)} \overline{\phi}(G_e)\tphi(G_e),\\
\la \psi |\tpsi\ra_{\SU(2)} &=&\int_{\SU(2)} \overline{\phi}(g_e)\tphi(g_e). \nn
\ees
These two evaluations are a priori very different. This can be seen also from the scalar product between the basis states \Ref{scalarproj} and \Ref{scalarsu}:
$$
\la \vphi_{\cI_e,j_e^{s,t},i_v}| \vphi_{\tilde{\cI}_e,\tilde{j}_e^{s,t},\tilde{i}_v}\ra
\,=\,
\prod_e\f{\delta_{n_e,\tilde{n}_e}\delta(\rho_e-\trho_e)}
{\mu(n_e,\rho_e)}
\,\delta_{j_e^{s,t},\tilde{j}_e^{s,t}}
\,\prod_v \la i_v|\tilde{i}_v\ra,
\qquad
\la\psi_{j_e,i_v}|\psi_{\tj_e,\tilde{i}_v}\ra_{\SU(2)}
\,=\,
\prod_e\f{\delta_{j_e,\tilde{j}_e}}{d_{j_e}}
\,
\,\prod_v \la i_v|\tilde{i}_v\ra,
$$
which differ in their measure and normalization. The key difference is due to the extra $\delta$-functions due to the $\SL(2,\C)$-irrep label, more specifically $\delta(\rho_e-\trho_e)$ which potentially could lead to divergences.

To illustrate this, we start with two $\SU(2)$ cylindrical functions $\psi,\tpsi$ and respectively apply the lifts $\cL_{\{\beta_e\}}$ and $\cL_{\{\tbeta_e\}}$. Then a straightforward calculation leads to:
\bes
\la\cL_{\{\beta_e\}}\psi|\cL_{\{\tbeta_e\}}\tpsi\ra
&=&
\int_{\SU(2)} [dk_e d\tk_e]\,
\overline{\psi}(k_e)\tpsi(\tk_e)\,
\prod_e \sum_{j_e} \f{\Delta_{j_e}^2}{d_{j_e}^2}\f{\delta(\rho_e-\trho_e)}{(\rho_e^2+j_e^2)}\chi^{j_e}(k_e^{-1}\tk_e) \nn\\
&=&
\prod_e\delta(\beta_e-\tbeta_e)\,
\int_{\SU(2)} [dk_e d\tk_e]\,
\overline{\psi}(k_e)\tpsi(\tk_e)\,
\prod_e \sum_{j_e} \f{\Delta_{j_e}^2}{d_{j_e}^2(j_e+1)(\beta_e^2(j_e+1)^2+j_e^2)}\chi^{j_e}(k_e^{-1}\tk_e).
\ees
Assuming the standard definition $\Delta_{j_e}=d_{j_e}^2$ ensuring that the lifts $\cL_{\{\beta_e\}}$ correctly invert the projection map $\cM$, then it is clear that the two scalar products do not match.
Then the natural question is which scalar product (between the Lorentz scalar product and the $\SU(2)$ scalar product) should we use on our kinematical Hilbert space of boundary states? This question should ultimately not matter so much since the final physical scalar should a priori be neither of them. Nevertheless, it is a crucial issue when building spinfoam amplitudes.

\medskip

An alternative would be to give up the requirement that a lift should be the inverse of the projection map $\cM$, i.e give up the idea that the restriction of the projected cylindrical function to the $\SU(2)$ subgroup be equal to the original $\SU(2)$ cylindrical function. Then we can modify the definition of the weight $\Delta_{j_e}$ and choose the new renormalized value, which now depends on the value of the Immirzi parameter $\beta_e$:
\be
\Delta_{j_e}^{\beta_e}\,\equiv\,
d_{j_e}^2\,\sqrt{(j_e+1)(\beta_e^2(j_e+1)^2+j_e^2)}.
\ee
This would define modified lifting maps, which would still send $\SU(2)$ cylindrical functions onto projected cylindrical functions, but that would conserve scalar products. Indeed, explicitly defining the new maps,
\be
\begin{array}{llcl}
L_{\{\beta_e\}}:
\,&H_S&\rightarrow&H_p \\
&\psi(g_e)&\mapsto&
\displaystyle{\vphi(G_e,B_v)=
\int_{\SU(2)} [dh_edk_e]\,
\psi(k_e)\,\sum_{j_e}\Delta_{j_e}^{\beta_e}\,
\chi^{j_e}(h_ek_e)\,\Theta^{(j_e,\beta_e(j_e+1))}(B_{s(e)}^{-1}G_eB_{t(e)}h_e).}
\end{array}\,,
\ee
using the new definition of the weight $\Delta_{j_e}^{\beta_e}$ given above, we will have the exact equality:
\be
\la L_{\{\beta_e\}}\psi|L_{\{\tbeta_e\}}\tpsi\ra
\,=\,
\la \psi|\tpsi\ra_{\SU(2)}\,\prod_e \delta(\beta_e-\tbeta_e).
\ee
Let us insist on the fact that this lifting map will still send the basis of $\SU(2)$ spin networks on projected spin network states satisfying the EPRL-FK ansatz, but with a different normalization that the lifting maps $\cL_{\{\beta_e\}}$ inverting $\cM$.

Finally, the natural issue is which lifting maps should we use to send LQG's $\SU(2)$ cylindrical functions onto the projected cylindrical functions of spinfoam models: should we enforce the matching condition that the restriction of projected cylindrical function to the $\SU(2)$ subgroup be equal to  the $\SU(2)$ cylindrical function or should we simply require the matching of the two scalar products and the unitarity of the lifting?




\section*{Conclusion}

In this short paper, we have investigated the correspondence between the $\SU(2)$ spin network states of the canonical loop quantum gravity framework and the projected spin networks arising in spinfoam models. After a detailed review of projected cylindrical functions and projected spin networks, we have introduced the projection map from projected cylindrical functions down to $\SU(2)$ cylindrical functions. Reversely, we have studied the lifting maps allowing to inverse this projection map and raise $\SU(2)$ spin network to projected spin networks on $\SL(2,\C)$. We have obtained a whole family of such lifting maps, parameterized by the Immirzi parameter, or more precisely an Immirzi field (i.e one value of the Immirzi parameter for each edge of the graph on which is defined the spin network). This way, we established an isomorphism between the space of $\SU(2)$ spin networks and the space of proper projected cylindrical functions at fixed Immirzi parameter. We have also shown that allowing the Immirzi parameter to run through all possible real values, we sweep the whole space of proper projected cylindrical functions. Finally, we have analyzed the differences between the two scalar products respectively for $\SU(2)$ functionals and $\SL(2,\C)$ functionals, and we have explained how to modify the lifting maps so as to ensure that these two scalar products match exactly.

This work hints towards considering that most useful perspective would be to compare $\SU(2)$ spin networks to projected spin networks and not directly to $\SL(2,\C)$ spin networks as was done in recent work on bridging between the EPRL-FK spinfoam models and the canonical approach \cite{eprl_jerzy}.
Physically, $\SL(2,\C)$ spin networks erase all data about the time-normal field, which is actually instrumental in properly implementing  the simplicity constraints. Mathematically, both $\SU(2)$ spin networks and projected spin networks involve $\SU(2)$ intertwiners, which allows for a direct map between the two Hilbert spaces. Therefore, we propose to use consistently projected spin networks as boundary states for the EPRL-FK spinfoam models and we hope that the present work will be useful in order to consistently translate Loop Quantum Gravity's dynamics into spinfoam amplitudes.

We have focused on the Lorentzian case with the gauge group $\SL(2,\C)$, but all our procedure applies to the Euclidean case based on the gauge group $\Spin(4)$. For the curious reader, we give the detail of the action of the $\spin(4)$ algebra in the $\SU(2)$ basis in appendix. Problems arise in this case when considering non-trivial values of the Immirzi parameter and the lifting maps do not exactly inverse the projection map. This is because $\Spin(4)$-irreps are labeled by a couple of (half-)integers and not by a continuous label (such as $\rho$) as in the Lorentzian case. Seen from this angle, the Lorentzian case can actually be considered as simpler than the Euclidean case.

We would like to also point out that our projected cylindrical functions obtained through a lift of $\SU(2)$ spin networks look similar to the recently introduced ``holomorphic" spin network functionals introduced to study the semi-classical behavior of the EPRL-FK spin amplitudes \cite{claudio,claudio2}. We think that this is an issue worth studying in more details.

Finally, we hope that the relation between $\SU(2)$ spin networks and projected functionals which we uncovered will trigger more interest in studying the structure of the space of projected spin networks. More particularly, we would like to put emphasis on two issues. First, it would be interesting to understand the geometrical interpretation of un-proper projected spin networks i.e states carrying two different spins  per edge $j_e^s\ne j_e^t$ (when the spin along an edge is different at its source vertex and at its target vertex). Then, it would be interesting to investigate the coarse-graining of projected cylindrical functions and see if we can construct a projective limit \`a la Ashtekar-Lewandowski as was done in Loop Quantum Gravity \cite{ALmeasure}. Such techniques have failed up to now when applied to spin network states for non-compact gauge groups such as the Lorentz group $\SL(2,\C)$. Nevertheless, we believe that this could be different when dealing with projected spin networks due to their effective $\SU(2)$ gauge invariance and their mapping into $\SU(2)$ spin networks.

\appendix

\section{Irreducible representations of $\Spin(4)$}

The group $\Spin(4)$ is isomorphic to the product of two subgroups, each isomorphic to $\SU(2)$: $\Spin(4)\sim \SU(2)_L\times \SU(2)_R$. Its algebra $\mathfrak{spin}(4)$ is the linear sum of two commuting algebras: $\mathfrak{su}(2)_L \oplus \mathfrak{su}(2)_R$. Its representations are labeled by two spins $(j_R,j_L)$. If we call $\vec{J}_R$ and $\vec{J}_R$ the standard generators of the left and right $\SU(2)$ groups, the generators of the space rotation group $\SU(2)$ are $\vec{J}=\vec{J}_L+\vec{J}_R$ while the ``boosts" generators are $\vec{K}=\vec{J}_R-\vec{J}_L$. The two Casimir operators are:
\be
C_1=\vJ^2+\vec{K}^2=2\vJ_L^2+2\vJ_R^2=2j_L(j_L+1)+2j_R(j_R+1),
\ee
and
\be
C_2= \vJ \cdot \vec{K}=\vJ_L^2-\vJ_R^2=j_L(j_L+1)-j_R(j_R+1)
\ee
We introduce
$$ J^{\pm}= J^1 \pm i J^2, \quad K^\pm = K^1 \pm i K^2;$$
the commutation relations defining the $\spin(4)$ algebra are then given by:
\bes
&&[J^+,J^3]=-J^+, \quad[J^-, J^3]=J^-, \quad [J^+, J^-]=2J^3 \nn \\
&& [J^+,K^+]=[J^-, K^-]= [J^3, K^3]=0, \quad [J^+, K^-]=-[J^-, K^+]=2K^3, \nn \\
&&[J^+, K^3]=-K^+, \quad [J^-,K^3]=K^-, \quad [K^+, J^3]=-K^+, \quad [K^-, J^3]=K^-, \\
&&[K^+,K^3]=-J^+, \quad[K^-, K^3]=J^-, \quad [K^+, K^-]=2J^3 \nn
\ees
Then for a given $\Spin(4)$ representation $(j_L, j_R)$, the action of the generators in the standard $\SU(2)$ basis noted as $|(j_L, j_R); j,m\ra$ or simply as $|j, m\ra$ is given by:
\bes
J^3\,|j, m\ra &=& m|j,m\ra, \nn \\
J^+\,|j,m\ra&=& \sqrt{(j-m)(j+m+1)}\,|j,m+1\ra, \nn \\
J^-\,|j,m\ra &=&\sqrt{(j+m)(j-m+1)}\,|j,m-1\ra, \nn \\
K^3\,|j,m\ra&=& \alpha_j \sqrt{j^2-m^2}\, |j-1,m\ra +\beta_jm \, |j,m\ra - \alpha_{j+1} \sqrt{(j+1)^2-m^2}\, |j+1,m\ra, \\
K^+\,|j,m\ra&=&\alpha_j \sqrt{(j-m)(j-m-1)}\, |j-1,m+1\ra +\beta_j \sqrt{(j-m)(j+m+1)} \, |j, m+1\ra \nn \\
&&  +\alpha_{j+1} \sqrt{(j+m+1)(j+m+2)}\,|j+1,m+1\ra, \nn \\
K^-\,|j,m\ra&= & -\alpha_j \sqrt{(j+m)(j+m-1)}\, |j-1,m-1\ra+\beta_j \sqrt{(j+m)(j-m+1)} \, |j,m-1 \ra \nn \\
&& -\alpha_{j+1} \sqrt{(j-m+1)(j-m+2)}\,|j+1,m-1\ra , \nn \\
\ees
where
\be
\beta_j=\f{(j_L+j_R+1)(j_L-j_R)}{j(j+1)},
\qquad
\alpha_j= \f{1}{j} \sqrt{\f{((j_L+j_R+1)^2-j^2)(j^2-(j_L-j_R)^2)}{4j^2-1}}.
\ee
As $\alpha_{j_L+j_R+1}=0$ and $\alpha_{|j_L-j_R|}=0$, the $\SU(2)$ spin $j$ runs from $|j_L-j_R|$ to $j_L+j_R$ as expected. An important particular case is for the so-called simple representations which are equivalently defined as the irreducible representations which contain a $\SU(2)$-invariant vector, or equivalently such that the second Casimir vanishes $C_2=\vJ \cdot \vec{K}=0$. Therefore, they are such that $j_L=j_R$ and we label them by a single half-integer $n=j_L=j_R$; then the action of the generators on $|n; j, m \ra$ remains the same but the coefficient $\alpha$ and $\beta$ are much simplier:
\be
\alpha_j^{(n)}= \sqrt{\f{C_1+1-j^2}{4j^2-1}}= \sqrt{\f{4n(n+1)-(j-1)(j+1)}{4j^2-1}}, \quad \textrm{ and } \beta_j^{(n)}=0.
\ee



\begin{thebibliography}{99}

\bibitem{clqg}
S. Alexandrov and D. Vassilevich,
{\it Area spectrum in Lorentz covariant loop gravity},
Phys.Rev. D64 (2001) 044023 [arXiv:gr-qc/0103105]

\bibitem{clqg0}
S. Alexandrov,
{\it Hilbert space structure of covariant loop quantum gravity},
Phys.Rev. D66 (2002) 024028 [S. Alexandrov]

\bibitem{clqg1}
S. Alexandrov and E.R. Livine,
{\it SU(2) Loop Quantum Gravity seen from Covariant Theory},
Phys.Rev. D67 (2003) 044009 [arXiv:gr-qc/0209105]

\bibitem{clqg2}
E.R. Livine,
{\it Towards a Covariant Loop Quantum Gravity},
arXiv:gr-qc/0608135,
chapter 14 in ''Approaches to quantum gravity", edited by D. Oriti, Cambridge University Press 2009

\bibitem{michael}
M. Reisenberger,
{\it A lattice worldsheet sum for 4-d Euclidean general relativity},
arXiv:gr-qc/9711052

\bibitem{projected}
E.R. Livine,
{\it Projected Spin Networks for Lorentz connection: Linking Spin Foams and Loop Gravity},
Class.Quant.Grav. 19 (2002) 5525-5542 [arXiv:gr-qc/0207084]


\bibitem{bcl}
J.W. Barrett and L. Crane,
{\it A Lorentzian Signature Model for Quantum General Relativity},
Class.Quant.Grav. 17 (2000) 3101-3118 [arXiv:gr-qc/9904025]

\bibitem{bcl_carlo}
A. Perez and C. Rovelli,
{\it Spin foam model for Lorentzian General Relativity},
Phys.Rev.D63 (2001) 041501 [arXiv:gr-qc/0009021]

\bibitem{bcl_normal}
E.R. Livine and D. Oriti,
{\it Coupling of spacetime atoms and spin foam renormalisation from group field theory},
JHEP0702 (2007) 092 [arXiv:gr-qc/0512002]

\bibitem{epr}
J. Engle, R. Pereira and C. Rovelli,
{\it The loop-quantum-gravity vertex-amplitude},
Phys.Rev.Lett.99 (2007) 161301 [arXiv:0705.2388]

\bibitem{eprl}
J. Engle, E.R. Livine, R. Pereira and C. Rovelli,
{\it LQG vertex with finite Immirzi parameter},
Nucl.Phys.B799 (2008) 136-149 [arXiv:0711.0146]

\bibitem{fk}
L. Freidel and K. Krasnov,
{\it A New Spin Foam Model for 4d Gravity},
Class.Quant.Grav.25 (2008) 125018 [arXiv:0708.1595]


\bibitem{ls2}
E.R. Livine and S. Speziale,
{\it Consistently Solving the Simplicity Constraints for Spinfoam Quantum Gravity},
Europhys.Lett.81 (2008) 50004 [arXiv:0708.1915]

\bibitem{roberto}
J.W. Barrett, R.J. Dowdall, W.J. Fairbairn, F. Hellmann and R. Pereira,
{\it Lorentzian spin foam amplitudes: graphical calculus and asymptotics},
Class. Quantum Grav. 27 (2010) 165009 [arXiv:0907.2440]


\bibitem{sf_valentin}
V. Bonzom and E.R. Livine,
{\it A Lagrangian approach to the Barrett-Crane spin foam model},
Phys.Rev.D79 (2009) 064034 [arXiv:0812.3456]

\bibitem{sf_daniele}
S. Gielen and D. Oriti,
{\it Classical general relativity as BF-Plebanski theory with linear constraints},
Class. Quantum Grav. 27 (2010) 185017 [arXiv:1004.5371]

\bibitem{sf_jimmy}
B. Dittrich and J. Ryan,
{\it Simplicity in simplicial phase space},
arXiv:1006.4295

\bibitem{sf_sergei1}
S. Alexandrov,
{\it Spin foam model from canonical quantization},
Phys.Rev.D77 (2008) 024009 [arXiv:0705.3892]

\bibitem{sf_sergei2}
S. Alexandrov,
{\it Simplicity and closure constraints in spin foam models of gravity},
Phys.Rev.D78 (2008) 044033 [arXiv:0802.3389]

\bibitem{eprl_jerzy}
W. Kaminski, M. Kisielowski and J. Lewandowski,
{\it The EPRL intertwiners and corrected partition function},
arXiv:0912.0540


\bibitem{noncompact}
L. Freidel and E.R. Livine,
{\it Spin Networks for Non-Compact Groups},
J.Math.Phys. 44 (2003) 1322-1356 [arXiv:hep-th/0205268]

\bibitem{eprl_ding}
Y. Ding and C. Rovelli,
{\it Physical boundary Hilbert space and volume operator in the Lorentzian new spin-foam theory},
arXiv:1006.1294


\bibitem{claudio}
E. Bianchi, E. Magliaro and C. Perini,
{\it Spinfoams in the holomorphic representation},
arXiv:1004.4550

\bibitem{claudio2}
E. Bianchi, E. Magliaro and C. Perini,
{\it Coherent spin-networks},
arXiv:0912.4054

\bibitem{ALmeasure}
A. Ashtekar and J. Lewandowski,
{\it Projective Techniques and Functional Integration},
J.Math.Phys. 36 (1995) 2170-2191 [arXiv:gr-qc/9411046]

\end{thebibliography}
\end{document}